\PassOptionsToPackage{dvipsnames}{xcolor}
\documentclass[nofootinbib, superscriptaddress]{revtex4-2}

\usepackage{amsmath}
\usepackage{amssymb}
\usepackage{float}
\usepackage{orcidlink}
\usepackage{subcaption}
\usepackage{url}
\usepackage{xcolor}

\newcommand*\dif{\mathop{}\!\mathrm{d}}

\begin{document}

\title{Numerical self-force calculations for scalar particles, formulated in the lab frame}

\author{Stamatis Vretinaris \orcidlink{0000-0001-7575-813X}}
\affiliation{Institute for Mathematics, Astrophysics and Particle Physics,
  Radboud University, Heyendaalseweg 135, 6525 AJ Nijmegen, The Netherlands}
\affiliation{Albert-Einstein-Institut, Max-Planck-Institut für Gravitationsphysik,
  Callinstraße 38, 30167 Hannover, Germany}
\affiliation{Leibniz Universität Hannover, 30167 Hannover, Germany}

\author{Erik Schnetter \orcidlink{0000-0002-4518-9017}}
\affiliation{Perimeter Institute for Theoretical Physics, Waterloo, Canada}
\affiliation{Department of Physics and Astronomy, University of Waterloo, Waterloo, Canada}
\affiliation{Center for Computation \& Technology, Louisiana State University, Baton Rouge, Louisiana, USA}

\author{Badri Krishnan \orcidlink{0000-0003-3015-234X}}
\affiliation{Institute for Mathematics, Astrophysics and Particle Physics,
  Radboud University, Heyendaalseweg 135, 6525 AJ Nijmegen, The Netherlands}
\affiliation{Albert-Einstein-Institut, Max-Planck-Institut für Gravitationsphysik,
  Callinstraße 38, 30167 Hannover, Germany}
\affiliation{Leibniz Universität Hannover, 30167 Hannover, Germany}

\date{2026-06-02}

\begin{abstract}
  We derive equations of motion for scalar particles self-consistently interacting with a scalar field, including the radiation produced by the particles' acceleration. Our approach differs in three key aspects from current methods: (1) we assume a small but finite discretization length scale $h$, which allows us to treat the particle as a small but finite object,  (2) we choose the state vector for the system before deriving equations of motion, and (3) we formulate the equations explicitly in the lab frame and not in a manifestly covariant manner. This approach, which is self-consistent, happens to greatly simplify the resulting equations and their derivation, and is directly suitable for numerical calculations. The result is an effective source method which generalizes to electrodynamics or general relativity in a straightforward manner (although we do not consider this here). We then provide two possible discretizations of these equations, based on finite volumes and spectral methods, and show results of one-dimensional calculations. These calculations show excellent agreement with analytic solutions.
\end{abstract}

\maketitle

\section{Introduction}

Finding solutions for physical systems which couple point particles and fields is difficult because point particles act as formally singular sources for the field. Doing so numerically is even more complicated. Yet such systems are physically of great interest because point particles are excellent approximations for small extended bodies. For example, in relativistic astrophysics, Extreme Mass Ratio Inspirals (EMRIs) are such system: There is a large black hole (with size $M$) and a secondary, much smaller compact object of size $m \ll M$. If the internal degrees of freedom of the small object can be neglected, then it is well approximated by a point particle.

While it is theoretically possible to discretize such a system using a length scale $h \ll m$ to resolve both bodies, the computational cost is prohibitive. For a standard discretization with a cutoff at wavelength $h$, the cost scales as
$O(1/h^4)$ in four-dimensional spacetime. Numerical calculations for such a system are thus intractable as the mass ratio $m/M$ decreases. Current state-of-the-art simulations find a mass ratio of $10^{-1}$ computationally feasible, and while $10^{-2}$ has been attempted, more extreme mass ratios are unfeasible with standard methods.
The gravitational self-force approach describes the motion of a small, compact body of mass $m \ll M$ as a perturbation on a fixed background spacetime. At leading order, the small body follows a geodesic of the background geometry; at first order, its own metric perturbation modifies that motion. The central difficulty is that the retarded field of a point mass diverges on its worldline, so any physically meaningful equation of motion requires a regularization prescription. Building on earlier radiation-reaction analyses in flat and curved spacetime \cite{Dirac1938,DeWitt1960}, Mino, Sasaki and Tanaka and Quinn and Wald derived the first-order gravitational self-force equations now known as the MiSaTaQuWa equations \cite{Mino1997,Quinn1997}. The Detweiler--Whiting decomposition then clarified the split of the retarded field into a singular piece, which exerts no force, and a regular piece that leads to acceleration \cite{Detweiler2003a}. Subsequent matched-asymptotic and self-consistent derivations established the gravitational equation of motion \cite{Gralla2008,Pound2010}, while extensions to second perturbative order developed the nonlinear gravitational self-force framework needed for phase-accurate EMRI waveforms \cite{Rosenthal2005,Rosenthal2006,Pound2012,Gralla2012,Detweiler2012}. For a comprehensive review of the self-force problem, see \cite{Poisson2011, Detweiler2005, Barack2009, Blanchet2011}.

To make the problem computationally tractable, one needs to define a finite regular field, or the force derived from it, without evaluating divergent quantities on the worldline. Dissipative and flux-based schemes avoid local regularization by using radiative fields or wave fluxes at infinity and the horizon, providing efficient adiabatic inspirals while omitting conservative effects \cite{Mino2003,Hughes2005,Drasco2006,Sundararajan2007}. For local self-force calculations, three regularization strategies are commonly used: Mode-sum regularization decomposes the retarded field into finite spherical-harmonic modes and subtracts analytic regularization parameters before summing over modes \cite{Barack2000,Barack2002}. Worldline-convolution methods compute the regular field from a Green function integral over the particle's past history, using quasilocal expansions together with frequency or time domain Green function calculations to make the approach practical \cite{Poisson1998,Anderson2005,Casals2013,Wardell2014,Zenginoglu2012}. Effective-source or puncture methods instead subtract an approximate singular field at the level of the differential equation, yielding a finite residual field suitable for time or frequency domain evolutions in $1+1$, $2+1$, or $3+1$ dimensions \cite{Barack2007,Vega2008,Dolan2011,Diener2012,Warburton2014}.

For the scalar self-force problem, self-consistent evolutions were first attempted by Diener et al.~\cite{Diener2012} using an effective-source method. In that implementation, however, numerical stability required omitting parts of the local singular-field approximation beyond velocity-dependent terms. Later, Heffernan et al.~\cite{Heffernan2018} extended the singular field to include acceleration-dependent contributions, but these terms introduce a self-referential dependence on higher time derivatives of the particle's self-acceleration. As discussed in Diener~\cite{Diener2020}, retaining such higher-derivative terms can lead to instabilities in the coupled particle-field evolution.

The first fully self-consistent evolutions of a scalar-charged particle on a Schwarzschild background were reported by Wittek et al.~\cite{Wittek2024,Wittek2025} using a worldtube excision framework. In their formulation, the acceleration-dependent part of the local analytical field introduces an implicit coupling: the puncture field depends on the particle's self-acceleration and its derivatives, while the self-acceleration is determined from the regularized field obtained through the matching procedure. At the order considered, evaluating the acceleration-dependent puncture and its time derivative requires quantities such as derivatives of the self-force, the first and second time derivatives of the four-velocity, and higher time derivatives of the regular field on the worldline. Rather than treating these higher derivatives as independent evolution variables or additional initial data, Wittek et al. close the system through an iterative scheme, initialized with the geodesic acceleration, in which the puncture, regular-field coefficients, and particle acceleration are updated successively. They further note that the acceleration terms can lead to instabilities in parts of parameter space, possibly reflecting the fact that the coupled point-particle system becomes effectively higher than second order in time; their iteration can therefore be viewed as analogous to a reduction-of-order treatment.

From the perspective of an initial-value formulation, the appearance of self-referential higher derivatives of the particle's acceleration is problematic. A well-posed evolution scheme should determine the future state of the coupled particle--field system from appropriate initial data for the field, the particle position, and the particle velocity. If the local singular-field approximation entering the effective source depends on time derivatives of the self-acceleration, then the force law is no longer a standard second-order equation of motion coupled to a hyperbolic field equation. Rather, the acceleration required to advance the worldline depends implicitly on higher derivatives of that same acceleration.

In this work, we develop a new framework for self-consistent scalar self-force evolutions that incorporates acceleration effects while preserving the structure of an initial-value formulation. The state of the coupled particle-field system is specified by a finite set of initial data, and its subsequent evolution is determined by equations that do not require higher time derivatives of the particle's acceleration. In contrast to formulations in which acceleration-dependent singular-field terms generate self-referential higher-derivative contributions, our approach avoids introducing additional derivative data beyond those required by the physical initial-value problem.

The framework is closely related in spirit to the effective-source method, but it does not rely on an expansion in powers of the particle's motion to construct the regularized evolution system. As a result, acceleration effects are incorporated naturally into the dynamics of the particle, rather than entering through higher-order local time-derivative corrections. This yields a self-consistent evolution scheme capable of accounting for accelerated motion while remaining compatible with a stable initial-value formulation.

Below, after giving details of the physical system we consider in section \ref{sec:system}, we derive the equations of motion in stages: we first consider a scalar field with a given charge distribution, then a particle moving in a given background field, and then we couple the two systems in section \ref{sec:eom}. We pay particular attention to explicitly factoring out a \emph{particle field} in section \ref{sec:particle-field}, as is common in self-force approaches. (Since we consider a finite-sized particle, this particle field is non-singular.) Finally, we describe our numerical methods in section \ref{sec:numerical-methods} and present results in section \ref{sec:numerical-results}.

\section{The physical system}
\label{sec:system}

\subsection{Overview}

We adopt a framework based on the hierarchy of scales $M \gg h
\gg m$ with $h$ being the grid scale considered.  We would like to keep $h$ larger than $m$ to make the problem computationally feasible.  Thus physics at scales smaller
than $h$, short-wavelength and high-frequency phenomena -- such as the
quasi-normal modes of the secondary -- are neglected. We approximate
the small black hole by a particle of effective size $\ell \lesssim
  h$. Within this approximation, $\ell$ is a free parameter independent
of $m$, provided it remains significantly smaller than the grid scale
$h$, so that the internal structure of the secondary can be neglected.

To simplify the present exposition, we forgo the complexities of
general relativity and consider a scalar particle coupled to a scalar
field in flat spacetime.

\subsection{A general remark on point particles}

Before describing our ansatz, we comment on the notion of a
\emph{point particle}. Point particles were probably first introduced
in a self-consistent manner by Newton, who showed that the
gravitational field of a spherically symmetric mass distribution is
equal to that of a point mass. This statement also holds in
electrodynamics and in general relativity. Point particles have been a
hugely successful concept in physics.

However, none of the currently known particles have been shown to be a
point particle by observational evidence, nor will this ever be
possible. Any observation and any experiment have a finite accuracy,
and we will thus only ever be able to put an upper limit on a
particle's size, and will never be able to show that a particle is
actually a point particle. This means that we use the concept of a
point particle only because it is mathematically convenient, not
because it is necessary for consistency with physics. For example, the
current upper limit for the electron radius is about 1~attometer, and
thus any calculation that assumes a finite electron radius smaller
than that will produce results that are consistent with all known
observations.

For simplicity, we will forgo the notion of a point particle here, and
instead consider spherically symmetric particles with a small but
finite radius $\ell$.

Finite-sized rigid particles are in principle incompatible with
special relativity. A force acting on one end of the particle would
need to accelerate that side of the particle before information about
the force has reached the other side, and a particle with internal
degrees of freedom (e.g. compressibility) seems the only option.
However, since we only study the spacetime with a finite resolution
$h$, we cannot see the details of such an interaction which happens at
a length scale $\ell \lesssim h$. Such approximations are ubiquitous
in numerical methods: Consider e.g. a finite volume method where the
manifold is split into cells of size $h$, and where the fluid in each
cell is described by a single, averaged velocity $v$. Such
discretizations are consistent, and in the same way a rigid particle
is consistent with relativity in our case. We discuss convergence in
the limit $h \to 0$ in section \ref{sec:numerical-methods} below.

\section{Equations of motion}
\label{sec:eom}

\subsection{A scalar field with a given charge distribution}

To set up notation we briefly describe the scalar field we consider
without particles present. The scalar field $\Phi$ is described by the
action
\begin{eqnarray}
  S &=& \frac{1}{2} \int  \eta^{ab} (\partial_a \Phi) (\partial_b \Phi)\,d^4x + \int \rho_\mathrm{e} \Phi\,d^4x
\end{eqnarray}
with an external (given) scalar charge distribution $\rho_\mathrm{e}$.  We assume
that $\rho_\mathrm{e}$ is given as a function of space and time. After
a $3+1$ split, the equations of motion are then
\begin{eqnarray}
  \label{eq:eom-basic}
  \partial_t \Phi &=& \Pi \\
  \partial_t \Pi &=& \delta^{ij} \partial_i \partial_j \Phi - \rho_\mathrm{e} \,\textrm{.}
\end{eqnarray}
Here we introduced a new variable $\Pi :=
  \partial_t \Phi$ to express the equations of motion as a system that
is first order in time. The state vector of the system, i.e. the
instantaneous conditions that define the future evolution of the
system, consists of instantaneous values for $\Phi$ and $\Pi$.

\subsection{A particle in a given scalar field}

Next we examine the motion of a scalar particle in a fixed background
field $\Phi_b$ which is given as a function of space and time. We
neglect the radiation emitted by the particle in this section.

We pay particular attention to defining the finite-sized particle in a
consistent manner. We define the particle's charge distribution in two
steps: First we consider a particle at rest at the origin, then a
particle moving with constant velocity.

\subsubsection{Particle at rest}

We define the charge distribution for a particle at rest at the origin
to be $\rho_0(x^i)$. As described above, we assume that $\rho_0$ has a
finite support of size $\ell$, with $\ell \lesssim h$ where $h$ is a
given length cutoff parameter.

This charge distribution is associated with a
scalar field $\Phi_0(x^i)$ through the Poisson equation
\begin{eqnarray}
  \delta^{ij} \partial_i \partial_j \Phi_0 &:=& \rho_0 \,\textrm{.}
\end{eqnarray}

This particle has a charge $Q$ and a mass (energy) $M$ given by
\begin{eqnarray}
  Q &:=& \int \rho_0\,d^4x
  \\
  M_0 &:=& m_0 + \frac{1}{2} \int \delta^{ij} (\partial_i \Phi_0) (\partial_j \Phi_0)\,d^4x + \int \rho_0 \Phi_0\,d^4x
  \\
  &=& m_0 + \frac{1}{2} \int \rho_0 \Phi_0\,d^4x \,\textrm{.}
\end{eqnarray}
Here, $m_0$ represents a possible intrinsic mass; we will set it to
zero below. Since we assume that the particle has a finite radius
$\ell$, all quantities above are well-defined and finite. These
conditions are \emph{constraints} on the possible choices of the
charge distribution function $\rho_0(x^i)$, if one wants to model a
particle with a given charge $Q$ and rest mass $M_0$.

\subsubsection{Moving particle}
\label{sec:moving-particle}

We consider a particle located at position $z^i$ and moving with a
constant three-velocity $v^i$. We describe the particle at a fixed
instance of time in the lab frame. This will allow us below to make
the particle part of the state vector; a state vector describes a
physical system at an instance of time.

The charge distribution and potential are given by boosting the
quantities $\rho_0$ and $\Phi_0$ defined above from the particle rest
frame to the lab frame:
\begin{eqnarray}
  \rho_\mathrm{lab}(z^i, v^i; x^i) &:=& \gamma \rho_0(\Lambda^i_j (x^i - z^i))
  \\
  \Phi_\mathrm{lab}(z^i, v^i; x^i) &:=& \Phi_0(\Lambda^i_j (x^i - z^i))
\end{eqnarray}
where $\Lambda^i_j = \Lambda^i_j(v^k)$ is the spatial part of the
Lorentz transformation for the velocity $v^i$, and $\gamma =
  \gamma(v^i) = 1/\sqrt{1 - v^2}$ is the respective Lorentz factor. We
treat $\rho_0$ as a scalar density that scales with $\gamma$ so that
the total particle charge $Q$ remains invariant under a Lorentz boost
($Q_\mathrm{lab} = Q$). A calculation shows that $M_\mathrm{lab} =
  \gamma M_0$ as expected.

\begin{quotation}
  We need to explain our notation to avoid confusion.
  $\rho_\mathrm{lab}$ and $\Phi_\mathrm{lab}$ are defined to be
  functions of three arguments: the particle position and velocity as
  well as a position in space. As defined, these functions do not have
  an explicit time dependence. These functions define, instantaneously
  in the lab frame, the charge distribution and field for a respective
  particle that moves uniformly.

  We will also be careful in our notation when using partial and total
  derivatives. A partial derivative is always applied to a function, and
  denotes the derivative with respect to one of the function's
  arguments. A total derivative is applied to an expression, and denotes
  the derivative with respect to one of the variables appearing in that
  expression. (This is the same distinction made e.g. in Mathematica
  between \verb+Derivative+ and \verb+D+.) For example, we can write
  $\partial/\partial x_i\, \rho_0$ (because $\rho_0$ is a function that
  depends on an argument named $x_i$), or we can write $d/dt\,
    \sin(t^2)$ (since the variable $t$ appears in the expression). It does
  \emph{not} make sense to write $\partial/\partial t\,
    \rho_\mathrm{lab}$ since $\rho_\mathrm{lab}$ does not have an argument
  named $t$. It \emph{does} make sense to write $d/dt\,
    \Phi_\mathrm{lab}(z^i + t v^i, v^i; x^i)$ because the variable $t$
  appears in that expression.
\end{quotation}

We will later also need the time derivatives of the particle field in
the lab frame. We define $\Pi_\mathrm{lab}$ and $\Xi_\mathrm{lab}$ to
be the first and second time derivatives of $\Phi_\mathrm{lab}$ that
would be seen in the lab frame for this non-accelerating particle:
\begin{eqnarray}
  \Pi_\mathrm{lab}(z^i, v^i; x^i) &:=& \frac{d}{dt} \Phi_\mathrm{lab}(z^i + t v^i, v^i; x^i) \\
  &=& v^i \partial_{z^i} \Phi_\mathrm{lab}(z^i, v^i; x^i) \\
  &=& - v^i \partial_i \Phi_\mathrm{lab}(z^i, v^i; x^i)
  \\
  \Xi_\mathrm{lab}(z^i, v^i; x^i) &:=& \frac{d}{dt} \Pi_\mathrm{lab}(z^i + t v^i, v^i; x^i) \\
  &=& v^i \partial_{z^i} \Pi_\mathrm{lab}(z^i, v^i; x^i) \\
  &=& - v^i \partial_i \Pi_\mathrm{lab}(z^i, v^i; x^i) \\
  &=& v^i v^j \partial_i \partial_j \Phi_\mathrm{lab}(z^i, v^i; x^i)
\end{eqnarray}
where we write $\partial_i$ for $\partial_{x^i}$.

With these definitions, the wave equation holds for the
non-accelerating particle:
\begin{eqnarray}
  - \Xi_\mathrm{lab} + \delta^{ij} \partial_i \partial_j \Phi_\mathrm{lab} &=& \rho_\mathrm{lab} \,\textrm{.}
\end{eqnarray}

All quantities defined here so far ($\rho_0$, $\Phi_0$,
$\rho_\mathrm{lab}$, $\Pi_\mathrm{lab}$, $\Xi_\mathrm{lab}$) are
defined without explicit time dependence. In other words, these are
functions that map a given particle state (position, velocity) to a
charge distribution and field \emph{instantaneously} in the lab frame.
This is an important point where this work differs from other
approaches which use radar coordinates, retarded Green's functions, or
similar methods that propagate information causally.

\subsubsection{Zeno's Arrow Paradox}
\label{sec:zeno}

In the arrow paradox \cite{Lee2015,Vlastos1966}, Zeno of Elea (c. 490–430 BC) argues:
\begin{itemize}
  \item[] … If everything when it occupies an equal space is at rest at that instant of time, and if that which is in locomotion is always occupying such a space at any moment, the flying arrow is therefore motionless at that instant of time and at the next instant of time but if both instants of time are taken as the same instant or continuous instant of time then it is in motion (Aristotle Physics, 239b5-7) \cite{Huggett2025}.

  \item[] … What is in motion moves neither in the place it is nor in one in which it is not (Diogenes Laertius, Lives of Famous Philosophers, ix.72) \cite{Laertius1925}.
\end{itemize}
In Zeno's arrow paradox, the arrow is momentarily at rest for each time instance. The paradox arises from the attempt to reconcile the discrete nature of time (as a sequence of instants) with the continuous nature of motion. This instantaneity is also present in numerical simulations at a given time step.
(Zeno's paradox is resolved for us because our state vector holds both the position and the velocity of the particle. Hence the particle is not ``at rest at that instant in time'' –– it retains its velocity, even if only a single instance of time is considered.)

Our definition of the particle field follows a similar philosophy as Zeno's paradox. We define it in the instantaneous rest frame of the particle. This could lead to confusion: since we assume uniform motion every time we define the particle field, one might be led to think that there is no acceleration, in the same way as in Zeno's paradox there is no motion. To better grasp the idea behind our definition of the particle field, one can imagine Zeno's arrow paradox, but instead of uniform motion, an arrow that is accelerating. At each time instance the arrow would not accelerate but move with constant velocity.
(This is not a paradox since the state vector does not contain the acceleration.)

\subsubsection{Particle equations of motion}

We describe the particle trajectory by the functions $z_\mathrm{p}^i(t)$ and
$v_\mathrm{p}^i(t)$. The well-known equations of motion for the particle in a
given external (background) potential $\Phi_\mathrm{e}$ are then
\begin{eqnarray}
  \label{eqn:particle-eom}
  \partial_t z_\mathrm{p}^i &=& v_\mathrm{p}^i
  \\
  \gamma \left( M_0 + Q \Phi_\mathrm{e} \right) \partial_t v_\mathrm{p}^i  &\approx& - Q \delta^{ij} \partial_j \Phi_\mathrm{e}(z_\mathrm{p}^i)
\end{eqnarray}
with the particle charge distribution
\begin{eqnarray}
  \rho_\mathrm{p}(t, x^i) &:=& \rho_\mathrm{lab}(z_\mathrm{p}^i(t), v_\mathrm{p}^i(t); x^i) \,\mathrm{.}
\end{eqnarray}
In this approximation, we assumed that the particle has size $\ell$
(i.e. that $\rho_\mathrm{p}$ has compact support of size $\ell$) and
that $\Phi_\mathrm{e}$ only varies slowly compared to $h \gg \ell$, so
that the curvature of $\Phi_\mathrm{e}$ can be neglected (i.e. that
the particle acceleration $a$ satisfies $a \ll 1/h$).

These equations of motion are equivalent to the covariant formulation \cite{Poisson2011}
\begin{eqnarray}
  \left( M_0 - Q \Phi_\mathrm{e} \right) \frac{\partial}{\partial\tau} u^a
  &=& - Q \left( \eta^{ab} + u^a u^b \right) \partial_b \Phi_\mathrm{e} \,\textrm{.}
\end{eqnarray}

We choose to keep the charge of the particle constant in time.

\subsection{Scalar field for a given particle trajectory}
\label{sec:particle-field}

We now discuss the evolution of a scalar field in the presence of a
scalar particle with a given trajectory. The emphasis in this section
lies on handling the field $\Phi$ with a given spatial resolution $h$
in the presence of a particle with a length scale $\ell$ where $\ell
  \lesssim h$. The particle's charge distribution introduces a very
small length scale $\ell$ that needs to be handled well to avoid a
large discretization error.

The particle trajectory (position and velocity) is given
as functions of time via $z_\mathrm{p}(t)$ and $v_\mathrm{p}(t)$, respectively.

Describing this system in a self-consistent manner is purely a matter
of choosing a suitable discretization method. As mentioned earlier, if
we were to choose a discretization length scale $h \ll \ell$, then
this would be straightforward. Our task here is to invent a
discretization method that is consistent and accurate for the case $h
  \gtrsim \ell$, i.e. a method which can handle a small particle and the
associated large field gradients near the particle.

To do so, we use the well-known approach to split the physical fields
$\Phi$ and $\Pi$ into two parts:
\begin{eqnarray}
  \label{eq:split}
  \Phi &=& \Phi_\mathrm{r} + \Phi_\mathrm{p} \\
  \Pi &=& \Pi_\mathrm{r} + \Pi_\mathrm{p}
\end{eqnarray}
where we define the \emph{particle fields} $\Phi_\mathrm{p}$ and $\Pi_\mathrm{p}$ in a
convenient manner so that they ``handle'' the large field gradients. The
\emph{remainder fields} $\Phi_\mathrm{r}$ and $\Pi_\mathrm{r}$ will then not have any
large gradients. Since we are introducing new degree of freedom here,
we are free to choose $\Phi_\mathrm{p}$ and $\Pi_\mathrm{p}$ (in principle) arbitrarily,
and this choice then defines $\Phi_\mathrm{r}$ and $\Pi_\mathrm{r}$.

It is important to note that neither $\Phi_\mathrm{r}$ nor $\Phi_\mathrm{p}$ can be
measured; only the total field $\Phi$ is physical. We can thus define
$\Phi_\mathrm{p}$ in any way we deem convenient. Neither $\Phi_\mathrm{r}$ nor $\Phi_\mathrm{p}$
must satisfy any causality conditions, only their sum $\Phi$ needs to
do so. In a manner of speaking, $\Phi_\mathrm{p}$ and $\Pi_\mathrm{p}$ can be viewed as
additional numerical basis functions that we introduce to keep
$\Phi_\mathrm{r}$ and $\Pi_\mathrm{r}$ smooth. This is an analogous idea to the Detweiler-Whiting split \cite{Detweiler2003a,Poisson2011, Barack2018, Vega2009,Heffernan2012, Detweiler2003b, Vega2011, Bourg2024, Heffernan2014,Wardell2012}.

\subsubsection{Representing the particle field}

We choose to define the particle field $\Phi_\mathrm{p}$ (at a time $t$) as the
field of a non-accelerating particle (at that same time $t$), as seen
in the lab frame. We already defined these quantities in section
\ref{sec:moving-particle} above:
\begin{eqnarray}
  \Phi_\mathrm{p}(t, x^i) &:=& \Phi_{\mathrm{lab}}(z_\mathrm{p}^i(t), v_\mathrm{p}^i(t); x^i) \\
  \Pi_\mathrm{p}(t, x^i) &:=& \Pi_{\mathrm{lab}}(z_\mathrm{p}^i(t), v_\mathrm{p}^i(t); x^i) \\
  \rho_\mathrm{p}(t, x^i) &:=& \rho_{\mathrm{lab}}(z_\mathrm{p}^i(t), v_\mathrm{p}^i(t); x^i) \,\textrm{.}
\end{eqnarray}
We remind the reader that $z^i_\mathrm{p}(t)$ and $v^i_\mathrm{p}(t)$ are functions of
time, while $z^i$ and $v^i$ are scalar values.

These definitions lead to the following time derivatives for $\Phi_\mathrm{p}$
and $\Pi_\mathrm{p}$:
\begin{eqnarray}
  \partial_t \Phi_\mathrm{p}(t, x^i) &=& \frac{d}{dt} \Phi_{\mathrm{lab}}(z_\mathrm{p}^i(t), v_\mathrm{p}^i(t); x^i) \\
  &=& (\partial_t z_\mathrm{p}^i)\, \partial_{z^i} \Phi_{\mathrm{lab}}(z_\mathrm{p}^i(t), v_\mathrm{p}^i(t); x^i)
  + (\partial_t v_\mathrm{p}^i)\, \partial_{v^i} \Phi_{\mathrm{lab}}(z_\mathrm{p}^i(t), v_\mathrm{p}^i(t); x^i) \\
  &=& \Pi_\mathrm{p}(t, x^i) + S_\Phi
  \\
  \partial_t \Pi_\mathrm{p}(t, x^i) &=& \frac{d}{dt} \Pi_{\mathrm{lab}}(z_\mathrm{p}^i(t), v_\mathrm{p}^i(t); x^i) \\
  &=& (\partial_t z_\mathrm{p}^i)\, \partial_{z^i} \Pi_{\mathrm{lab}}(z_\mathrm{p}^i(t), v_\mathrm{p}^i(t); x^i)
  + (\partial_t v_\mathrm{p}^i)\, \partial_{v^i} \Pi_{\mathrm{lab}}(z_\mathrm{p}^i(t), v_\mathrm{p}^i(t); x^i) \\
  &=& \Xi_\mathrm{p}(t, x^i) + S_\Pi \\
  &=& \delta^{ij} \partial_i \partial_j \Phi_\mathrm{p}(t, x^i) - \rho_\mathrm{p}(t, x^i) + S_\Pi(t, x^i)
\end{eqnarray}
with the effective source terms \cite{Vega2008, Vega2009, Vega2011, Wardell2012}
\begin{eqnarray}
  S_\Phi &:=& (\partial_t v_\mathrm{p}^i)\, \partial_{v^i} \Phi_{\mathrm{lab}}(z_\mathrm{p}^i(t), v_\mathrm{p}^i(t); x^i)
  \\
  S_\Pi &:=& (\partial_t v_\mathrm{p}^i)\, \partial_{v^i} \Pi_{\mathrm{lab}}(z_\mathrm{p}^i(t), v_\mathrm{p}^i(t); x^i) \,\textrm{.}
\end{eqnarray}

The source terms are proportional to the particle acceleration
$\partial_t v_\mathrm{p}$. The source terms are fully defined by the functions
$\Phi_\mathrm{lab}$ and $\Pi_\mathrm{lab}$ which we introduced above.
The source terms are known functions of the particle position,
velocity, and acceleration. They describe how the instantaneous field
of the particle ($\Phi_\mathrm{p}$ and $\Pi_\mathrm{p}$) changes when the particle's
velocity changes.

We note that the source terms depend only on the change of velocity $\partial_t v^i$ and not on the change of position $\partial_t z^i$. The reason is that the latter terms cancel because we intentionally chose the particle field to satisfy the Poisson equation in a moving reference frame. In principle, other choices for the particle field would be possible, and in that case, the position-dependent source terms might not cancel.

\subsubsection{Equations of motion}

The equations of motion for the remainder fields $\Phi_\mathrm{r}$ and $\Pi_\mathrm{r}$
are now automatically also defined. They follow from the equations of
motion for the physical fields $\Phi$ and $\Pi$. They are
\begin{eqnarray}
  \label{eq:eom}
  \partial_t \Phi_\mathrm{r} &=& \partial_t \Phi - \partial_t \Phi_\mathrm{p} \\
  &=& \Pi - \Pi_\mathrm{p} - S_\Phi \\
  &=& \Pi_\mathrm{r} - S_\Phi
  \\
  \partial_t \Pi_\mathrm{r} &=& \partial_t \Pi - \partial_t \Pi_\mathrm{p} \\
  &=& \delta^{ij} \partial_i \partial_j \Phi - \rho - \delta^{ij} \partial_i \partial_j \Phi_\mathrm{p} + \rho_\mathrm{p} - S_\Pi \\
  &=& \delta^{ij} \partial_i \partial_j \Phi_\mathrm{r} - S_\Pi
\end{eqnarray}

These are the same equations as for a scalar field without a particle,
but with added source terms.

While we can assume that $\Phi_\mathrm{r}$ and $\Pi_\mathrm{r}$ are smooth, i.e. have no
important features below the length scale $h$, this is not the case
for the source terms. The source terms contain nontrivial features at
small length scales down to $\ell$. It is therefore necessary to
filter or smoothen them in (\ref{eq:eom}) to avoid contaminatination
from small wavelength~/ high frequency noise.

\section{Numerical methods}
\label{sec:numerical-methods}

For simplicity, we study the scalar particle and its fields only in
one dimension, not in three dimensions. With ``one dimension'' we
refer to a truly 1+1-dimensional spacetime, not to a spherically
symmetric reduction of a 3+1-dimensional spacetime.

We consider two distinct numerical implementations for the field
equations: a finite volume approximation and a spectral approximation.
These methods offer different perspectives on the spatial
representation of the fields and the treatment of the particle source.

It is convenient to choose a particle size $\ell$ that is very small,
i.e. $\ell \ll h$. This simplifies evaluating or approximating the
expressions for the source terms, and we will do so below.

\subsection{Finite Volume Method}
\label{sec:fvm}

We use the Finite Volume Method (FVM) \cite{LeVeque2002, Godunov1959} for spatial
discretization due to the nature of the source terms. The variational
approach (weak solutions) of FVM is suitable for treating source terms
that are integrable but not continuous, such as our case. (Although
the source terms are continuous in our case, they have very large
gradients, and thus ``appear'' discontinuous at the discrete level.)
We follow the Method of Lines (MoL) approach \cite{Schiesser2012},
discretizing the spatial domain while keeping time continuous. This
results in a system of ordinary differential equations (ODEs) which we
evolve using a Strongly Stability Preserving Runge-Kutta (SSPRK)
method \cite{Ruuth2006}.

Standard FVM methods have been designed for conservation laws where it
is important to handle discontinuities well. Our case is different --
we have source terms that have low regularity, i.e. which are ``almost
singular'' -- and we will therefore spend some time discussing the
fundamentals of FVM, justifying our choice.

In this subsection we will describe the FVM method in 1D for brevity.
The discussion would extend to higher dimensions in a straightforward
manner.

\subsubsection{First-order formulation}

We rewrite the system (\ref{eq:eom}) above in a fully first-order form
by introducing a new variable $\Psi_i := \partial_i \Phi$:
\begin{eqnarray}
  \partial_t \Phi_\mathrm{r} &=& \Pi_\mathrm{r} - S_\Phi \\
  \partial_t \Pi_\mathrm{r} &=& \delta^{ij} \partial_i \Psi_{\mathrm{r},j} - S_\Pi \\
  \partial_t \Psi_{\mathrm{r},i} &=& \partial_i \Pi_\mathrm{r} - S_\Psi
\end{eqnarray}
where
\begin{eqnarray}
  S_\Psi &:=& \partial_i S_\Phi \,\textrm{.}
\end{eqnarray}

\subsubsection{Characteristics}

We represent the field equations as a first-order hyperbolic system.
We define the field state-vector $U_f$ and the evolution system as:
\begin{eqnarray}
  \partial_t \bar{U}_f + A \partial_x \bar{U}_f = S
  \label{eq:first-order-system}
\end{eqnarray}
where
\begin{equation}
  U_f = \begin{pmatrix}
    \Pi_\mathrm{r}  \\
    \Psi_\mathrm{r} \\
    \Phi_\mathrm{r}
  \end{pmatrix}, \quad A=\begin{pmatrix}
    0  & -1 & 0 \\
    -1 & 0  & 0 \\
    0  & 0  & 0
  \end{pmatrix}, \quad S=\begin{pmatrix}
    - a^i \partial_{v^i} \Pi_\mathrm{lab}  \\
    - a^i \partial_{v^i} \Psi_\mathrm{lab} \\
    - a^i \partial_{v^i} \Phi_\mathrm{lab}
  \end{pmatrix},
\end{equation}
where $U_f$ is the field part of the state-vector, $A$ is the
principal part and $S$ is the source.

The principal part $A$ admits an eigendecomposition
\begin{equation}
  A = R \Lambda R^{-1}
\end{equation}
where $R$ contains the right eigenvectors $r^m$
\begin{equation}
  R = [r^1|r^2|...|r^m]
\end{equation}
and the matrix $\Lambda$ contains its eigenvalues $\lambda^m$
\begin{equation}
  \Lambda = \mathrm{diag}(\lambda^1, \lambda^2,...,\lambda^m).
\end{equation}

\subsubsection{Computational grid}
\label{sec:computational-grid}

We discretize the spatial domain $[-L,L]$ into $N$ cells of width $h$.
We employ a staggered grid where the $i$-th cell is defined by the
interval
\begin{equation}
  \Omega_i = (x_{i-1/2}, x_{i+1/2}).
\end{equation}
where $x_{i\pm1/2}=x_i \pm h/2$. The discrete variable $\bar{U}_i^n$
describes the average value
\begin{equation}
  \bar{U}_i^n = \frac{1}{h}\int_{\Omega_i} U(t_n, x) \dif x.
\end{equation}
By working with cell
averages we can ensure the numerical method is conservative even in
the presence of discontinuities. Averaging is
equivalent to computing the coefficients of an $L^2$
orthogonal projection onto the space of piecewise constant functions.

The spatial discretization of the field $U(\cdot, x)$ can be described as
projection onto a subspace of piecewise constant
functions. Let the domain be partitioned into disjoint cells
$\Omega_i$ of width $h$. We define the computational basis as the set
of characteristic functions $\{ b_i(x) \}$ with
\begin{equation}
  b_i(x) =
  \begin{cases}
    1 & \text{if } x \in \Omega_i, \\
    0 & \text{otherwise}.
  \end{cases}
\end{equation}
These functions form an orthogonal basis for the subspace $V_h \subset
  L^2(\mathbb{R})$. The approximation of the field, denoted $U_h \in
  V_h$, is constructed via the orthogonal projection operator
$\mathcal{P}_h: L^2 \to V_h$. This is defined by the condition that
the projection error is orthogonal to the subspace,
\begin{equation}
  \langle U - \mathcal{P}_h U, \, b_i \rangle = 0 \quad \forall i,
\end{equation}
where $\langle f, g \rangle = \int f(x) g(x) \dif x$ denotes the
standard inner product. Expanding the projection in terms of the
basis, $\mathcal{P}_h U(\cdot, x) = \sum_j \bar{U}_j b_j(x)$, and
substituting into the orthogonality condition yields
\begin{equation}
  \sum_j \bar{U}_j \langle b_j, b_i \rangle = \langle U, b_i \rangle.
\end{equation}
Due to the disjoint support of the basis functions,
the expansion coefficients $\bar{U}_j$ are
determined as
\begin{equation}
  \label{eqn:fv-projection}
  \bar{U}_j = \frac{1}{h} \int_{\Omega_j} U(x) \dif x.
\end{equation}
The standard cell-averaging operation is equivalent to the
coefficient extraction for the $L^2$ projection of the continuum field
onto the piecewise constant basis functions.

The above is relevant because our source terms are not regular and need to be properly projected into our function space. For smooth source functions, it is often possible (and often done) to evaluate them at the centre of the cells as approximation of this integral. This approach does not work here because the respective approximation error would be unbounded. Instead, (\ref{eqn:fv-projection}) needs to be used for the source term, as we describe in more detail below.

\subsubsection{Weak formulation}

Since \eqref{eq:first-order-system} contains low regularity source
terms, we interpret the evolution in a weak sense. Specifically, we
require the residual of the evolution equation to be orthogonal to our
basis functions $b_i(x)$:
\begin{equation}
  \left\langle \partial_t U + \partial_x (AU) - S, \, b_i \right\rangle = 0.
\end{equation}
Expanding the inner product using the definition of $b_i(x)$ yields:
\begin{equation}
  \int_{\Omega_i} \partial_t U \dif x + \int_{\Omega_i} \partial_x (AU) \dif x = \int_{\Omega_i} S \dif x.
\end{equation}
Applying the fundamental theorem of calculus to the flux term and
substituting the cell average $\bar{U}_i$, we obtain the semi-discrete
scheme:
\begin{equation}
  h \frac{d\bar{U}_i}{dt} + \left[ AU(t, x_{i+1/2}) - AU(t, x_{i-1/2}) \right] = \int_{\Omega_i} S \dif x.
\end{equation}
By defining the physical flux $F_{i\pm1/2} = AU(t, x_{i\pm1/2})$ and
the cell-averaged source $S_i = \frac{1}{h} \int_{\Omega_i} S \dif x$,
we arrive at the standard FVM form:
\begin{equation}
  \frac{d\bar{U}_i}{dt} = -\frac{1}{h} (F_{i+1/2} - F_{i-1/2}) + S_i.
  \label{eq:FVM_final}
\end{equation}
This formulation demonstrates that the FVM is mathematically
equivalent to a weighted residual method \cite{Finlayson2013}, where the test functions are
the same piecewise constant functions used for the $L^2$ projection.
This choice is also known as a Bubnov-Galerkin method \cite{Bubnov1913, Galerkin1915}. For higher order schemes where the reconstruction is done with linear interpolation, instead of piecewise constant, the trial functions are now piecewise linear while the test functions remain piecewise constant. This is now called a Petrov-Galerkin method \cite{Petrov1940}. Note that the
equation is in semi-discrete form, suitable for the MoL method.

\subsubsection{Numerical fluxes}

Since the flux calculations are being performed at the cell faces, we
need to introduce a numerical flux to approximate the fluxes at the
faces based on the values of $U$ at the cell centers. We use the
Godunov flux \cite{Godunov1959}, which is the same as the Rusanov flux
\cite{Rusanov1962} for the linear wave equation, since the local
Lax-Friedrichs speed \cite{Friedrichs1971} is given  exactly by the eigenvalues of the system,
\begin{equation}
  F^n_{i-1/2} = A^+ U_{i-1}^{\mathrm{R}} + A^- U_i^{\mathrm{L}}
\end{equation}
where $A^+$ and $A^-$ are defined as
\begin{equation}
  A^+ = R \Lambda^+ R^{-1} \, \textrm{and} \, A^- = R \Lambda^- R^{-1}
\end{equation}
where $\Lambda^\pm$ are the diagonal matrices containing only the
positive or negative eigenvalues respectively. Note that the values
$U_{i-1}^{\mathrm{R}}$ and $U_i^{\mathrm{L}}$ are evaluated on the
common face at $x_{i-1/2}$ between the cells $\Omega_{i-1}$ and
$\Omega_i$. As the state vector quantities ``live'' in the cell
centers, we have to reconstruct the values at the faces using
conservative interpolation. First order schemes use piecewise constant
interpolation, while for second order schemes, piecewise linear
interpolation is needed.

\subsubsection{Second-order approximation}

To resolve discontinuous data at second order, a standard practice is
to employ slope limiters and linear reconstruction in characteristic
space. These methods are also known as Monotonic Upstream-centered
Scheme for Conservation Laws (MUSCL) \cite{VanLeer1977,VanLeer1979}. The
homogeneous part is thus ensured to be Total Variation Diminishing
(TVD) \cite{Harten1983}. It is important to use a \emph{monotone}
flux \cite{LeVeque2002}, such as the Godunov or Rusanov, in order to
be compatible with the MoL.

\subsubsection{Source treatment}

We employ an unsplit method for the time evolution of the field
equations, wherein the source term is incorporated directly into the
right-hand side of the homogeneous evolution equations. The source
contribution is computed analytically by evaluating the cell-averaging
integral over each computational cell. To accurately capture the
behavior at the particle location, we split the integral of the cell that contains the particle at the particle location. This places the particle at an integration boundary, which simplifies evaluating the integral:
\begin{equation}
  S_i = \frac{1}{h} \left( \int_{x_{i-1/2}}^{z_\mathrm{p}} S_{\text{left}} \dif x + \int_{z_\mathrm{p}}^{x_{i+1/2}} S_{\text{right}} \dif x \right).
\end{equation}

\subsubsection{Convergence analysis}

While smooth solutions allow for
pointwise convergence analysis in the $L_\infty$ (max) norm,
discontinuous data typically undergoes ``smearing'' over a few grid
cells due to numerical diffusion. In such cases, the error at the jump
does not decrease in the infinity norm regardless of the grid refinement.
Therefore, we quantify convergence using the $L_1$ norm, which
represents the integral of the absolute error (see 8.1.1 from \cite{LeVeque2002}). To calculate the
convergence order $p$ without an analytical solution, we compare three
successive resolutions ($h, h/2, h/4$):
\begin{equation}
  p = \log_2 \frac{\| \bar{U}_{h} - \mathcal{R} \bar{U}_{h/2} \|_1}{\| \mathcal{R} \bar{U}_{h/2} - \mathcal{R}\mathcal{R} \bar{U}_{h/4} \|_1},
  \label{eq:self-convergence}
\end{equation}
where $\mathcal{R}$ denotes a restriction operator that averages the
higher-resolution cells onto the coarser grid. This averaging ensures
the comparison is conservative and consistent with the finite volume
projection.

The convergence behavior for non-smooth data is fundamentally limited.
As shown by LeVeque \cite{LeVeque2002} (pg 156), for a first-order
upwind method applied to the scalar advection of a Heaviside function, the
$L_1$ error behaves as $\mathcal{O}(h^{1/2})$. This is because
the numerical width of the discontinuity—the region where the solution is
smeared—typically grows as $\sqrt{t h}$. Consequently, even if
the scheme is
formally first-order for smooth data, the presence of a jump
discontinuity causes the $L_1$ error to decay only as $\sqrt{h}$.

By using the $L_1$ norm and a second-order MUSCL scheme
\cite{VanLeer1977,VanLeer1979}, we aim to recover a convergence rate closer to
$\mathcal{O}(h^1)$ for discontinuous solutions, but we
acknowledge that the $\Delta h^{1/2}$ limit provides a theoretical
baseline for the ``worst-case'' performance near the particle.

\subsection{Spectral method}

We also discretize the equations using a spectral method. We do this
for two reasons: First, spectral methods are perceived to very
``mathematical'' and ``difficult to get right''. By show that spectral
methods are working we hope to increase acceptance of our approach in
the community. And second, spectral methods are very simple to
implement for a linear system, and it is straightforward to apply a
filter to the source terms.

We choose a one-dimensional domain of size $[-L, +L]$ with $L=1$. For
simplicity we use Dirichlet boundary conditions where we set
$\Phi_\mathrm{r}=0$ and $\Pi_\mathrm{r}=0$. We restrict our evolution to the time
range $[0, 0.4L]$ and only analyze the resuls in the inverval $[-0.5L,
      +0.5L]$ so that information from the boundary conditions does not
influence our results.

We expand $\Phi_\mathrm{r}$ and $\Pi_\mathrm{r}$ into Fourier modes. Given the boundary
conditions, all $\cos$ terms vanish, and our basis functions are
\begin{eqnarray}
  b^n(x) &:=& \sin \pi n \frac{x+1}{2}
\end{eqnarray}
for $n \in 1 \ldots N$. These basis function automatically impose the
boundary conditions.

In this basis, the Laplace operator is diagonal and is given by
\begin{eqnarray}
  \Delta &:=& \mathrm{diag}\left[ - \left( n \frac{\pi}{2} \right)^2 \right]
\end{eqnarray}

When projecting the source terms onto our bases functions we use a
Gauss-Legendre quadrature \cite{FastGaussQuadrature2025} with $3200$
nodes. These are many more nodes than basis functions, and we assume
that our numerical quadrature is essentially exact up to
floating-point round-off error for our algorithm.

Our state vector consists of
\begin{eqnarray}
  \Phi_\mathrm{r}^n &:& \textrm{Coefficients for remainder field} \\
  \Pi_\mathrm{r}^n &:& \textrm{Coefficients for time derivative of remainder field} \\
  z_\mathrm{p} &:&: \textrm{Particle position} \\
  v_\mathrm{p} &:& \textrm{Particle velocity}
\end{eqnarray}

To evolve the coupled particle/field system we use the method of lines
(MoL) (see \ref{sec:fvm}). We evaluate the right hand side (RHS) of
the state vector in two steps:
\begin{enumerate}
  \item We evaluate the fields $\Phi_\mathrm{r}$ and $\Pi_\mathrm{r}$ and their gradients
        at the particle position $z$. We do this exactly, summing the
        spectral basis functions and their derivatives. We then evaluate the
        terms in the particle's equations of motion
        (\ref{eqn:particle-eom}). We also add a forcing term for some test
        cases (see below).
  \item We apply the Laplace operator to $\Phi_\mathrm{r}^n$. Since we already
        calculated the particle acceleration $a = d/dt\, v$ we can evaluate
        the source terms at each Gauss-Legendre quadrature node. This allows
        us to explicitly project the source terms onto each basis function,
        filtering out all high-frequency noise.
\end{enumerate}

It well known that specral methods converge ``spectrally''
(essentially exponentially) for smooth functions, i.e. that the error
decays as $\exp( -\alpha N)$ for $N$ spectral collocation points. The
clause ``for smooth functions'' does the heavy lifting in this
statement. In our case, the source terms are not smooth in this sense,
and we thus only expect first order convergence, i.e. an error
proportional to $1/N$. (A Fourier expansion converges for a large
class of functions, including discontinuous functions, and the
coefficients fall of as $1/N$.)

There exist well-known methods to improve the accuracy and convergence
order in the presence of non-smooth term. In one dimension one can
split the domain into two, with a moving domain boundary at the
particle location; this would restore the spectral convergence.
Similar approaches are possible in higher dimensions. We do not pursue
these approaches here to keep our presentation simple.

\section{Numerical results}
\label{sec:numerical-results}

For simplicity we consider a scalar field in $1+1$ dimensions. While
this greatly simplifies the numerical implementation, it is also more
difficult to understand intuitively because we are trained to think of
a particle as having a $1/r$ potential. This is not the case in $1+1$
dimensions where the potential is proportional to $|x|$.

To allow for a nontrivial scenario we consider a single scalar
particle which, in addition to the scalar field, also sees a static
harmonic potential. This exerts a periodic external force onto the
particle where the particle radiates and loses energy, leading to a
shrinking oscillation amplitude.

\subsection{Finite Volume Method results}

For the numerical implementation, we employ a five-stage,
fourth-order SSPRK \cite{Ruuth2006} scheme for time integration.
Spatial discretization is achieved via a second-order MUSCL
reconstruction \cite{VanLeer1977,VanLeer1979} using the Monotonized Central (MC)
slope limiter \cite{VanLeer1977} in characteristic space.

We validate our framework by simulating a harmonically
oscillating particle, evaluating the convergence of the numerical
solution against both the analytically calculated retarded field and its own
self-convergence.
We forgo explicit
boundary conditions for the scalar field; instead, we allow the field
to evolve freely and truncate the computational domain to exclude any
regions contaminated by numerical artifacts from the boundaries.

\subsubsection{Forced motion}

We simulate a particle with charge $q=1$ that oscillates with
maximum velocity $v_{max}=0.1$ and amplitude $A=0.01$. We evolve for a
total time $t_f=2.5$ during which the particle oscillate 4 times, thus we choose the
domain $[-L,L]$ with $L=6$. This domain size is chosen specifically so that, during the time frame of the four oscillations, any numerical artifacts or ``junk'' information from the boundaries cannot propagate into the region of interest. This allows us to safely truncate the domain as previously described, without loss of accuracy. Our baseline resolution is $dx_0=1$, which we refine while keeping a
constant Courant-Friedrichs-Lewy factor of $\texttt{cfl}=0.1$.

\begin{figure}[H]
  \begin{subfigure}[b]{0.48\textwidth}
    \includegraphics[width=\textwidth]{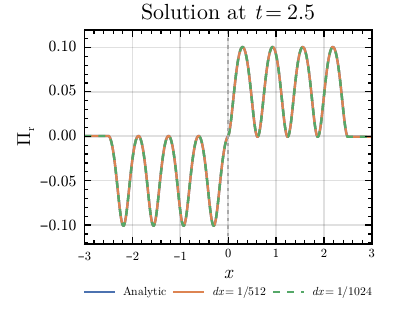}
    \label{fig:pifield-resolutions}
  \end{subfigure}
  \hfill
  \begin{subfigure}[b]{0.48\textwidth}
    \includegraphics[width=\textwidth]{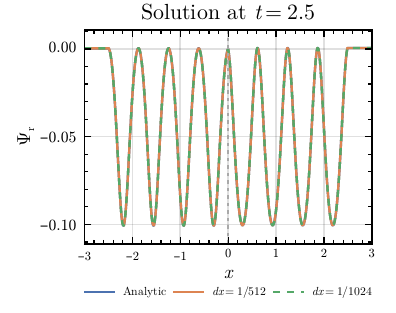}
    \label{fig:psifield-resolutions}
  \end{subfigure}
  \caption{Field $\Pi_\mathrm{r}$ (left) and $\Psi_\mathrm{r}$ (right)
    at a late time $t=2.5$ for two resolutions $dx=1/512$ and $dx=1/1024$, and comparing to the analytic
    retarded solution. The vertical dashed line indicates the
    position of the particle at that time.}
  \label{fig:field-resolutions}
\end{figure}

We find that our numerical fields show agreement with the
analytic retarded solutions for both $\Pi_\mathrm{r}$ and $\Psi_\mathrm{r}$, as demonstrated in Figure \ref{fig:field-resolutions}.

\begin{figure}[H]
  \begin{subfigure}[b]{0.48\textwidth}
    \includegraphics[width=\textwidth]{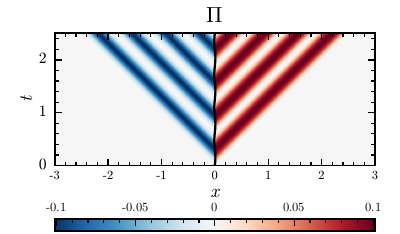}
  \end{subfigure}
  \hfill
  \begin{subfigure}[b]{0.48\textwidth}
    \includegraphics[width=\textwidth]{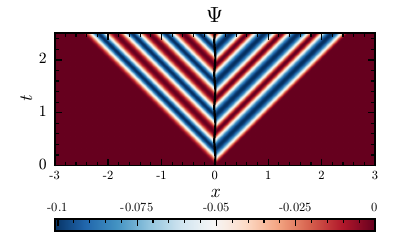}
  \end{subfigure}
  \caption{Space-time evolution of $\Pi_\mathrm{r}$ (left) and $\Psi_\mathrm{r}$ (right) for a particle in a harmonic potential with $q=1$ and $v_{max}=0.1$ for $N_{osc}=4$ periods for $h=1/512$. The field is initially zero everywhere, and the oscillating particle generates waves that propagate outwards. See Figure \ref{fig:spacetime-fields_near_particle} for a zoom onto the particle's trajectory.}
  \label{fig:spacetime-fields}
\end{figure}

\begin{figure}[H]
  \begin{subfigure}[b]{0.48\textwidth}
    \includegraphics[width=\textwidth]{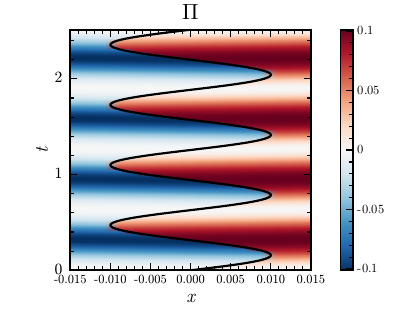}
    \label{fig:pifield-spacetime_near_particle}
  \end{subfigure}
  \hfill
  \begin{subfigure}[b]{0.48\textwidth}
    \includegraphics[width=\textwidth]{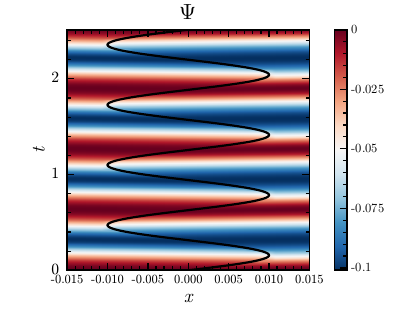}
    \label{fig:psifield-spacetime_near_particle}
  \end{subfigure}
  \vfill
  \caption{Zoom onto the particle: Space-time evolution of $\Pi_\mathrm{r}$ (left) and $\Psi_\mathrm{r}$ (right) for a particle in a harmonic potential with $q=1$ and $v_{max}=0.1$ for $N_{osc}=4$ periods for $h=1/1024$. The waves emanating from the oscillating particle are clearly visible.}
  \label{fig:spacetime-fields_near_particle}
\end{figure}

We also show the evolution of the fields in spacetime
Figure \ref{fig:spacetime-fields}, and zoom onto the particle in Figure \ref{fig:spacetime-fields_near_particle} to see the particle trajectory as well, for a resolution of $h=1/512$.

Since we can calculate the retarded field analytically, we can evaluate the convergence rate of our numerical scheme by comparing the computed fields against the analytic retarded fields throughout the entire evolution and investigate further at late times, such as $t=2.5$.

\begin{figure}[H]
  \begin{subfigure}[b]{.49\textwidth}
    \centering
    \includegraphics[width=\textwidth]{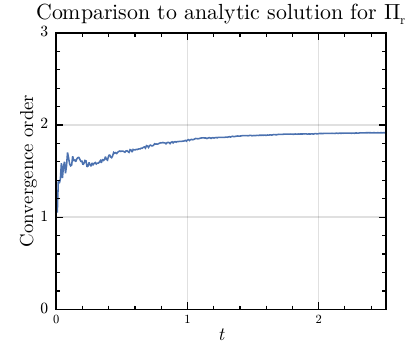}
  \end{subfigure}
  \hfill
  \begin{subfigure}[b]{.49\textwidth}
    \centering
    \includegraphics[width=\textwidth]{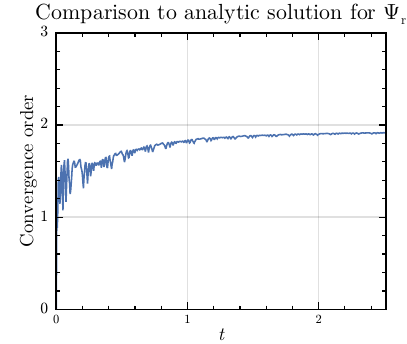}
  \end{subfigure}
  \caption{Convergence order for $\Pi_\mathrm{r}$ (left) and $\Psi_\mathrm{r}$ (right) fields against the analytic retarded fields, with resolutions $h=1/512$ and $h=1/1024$. As expected, our calculations are overall second order convergent. At early times the convergence order is reduced to about 1.5 for two reasons: (a) near the particle we only expect a convergence order of $1/2$ (as explained by LeVeque, see main text), and (b) the error is zero almost everywhere else in the domain because the radiation from the particle has not propagated yet. At late times, the small region near the particle does not affect the overall convergence much, as can be seen in Figure \ref{fig:field-scaled-errors-retarded}.
  }
  \label{fig:field-retarded-convergence-order}
\end{figure}

In Figure \ref{fig:field-retarded-convergence-order} we show the convergence order for $\Pi_\mathrm{r}$ (left) and $\Psi_\mathrm{r}$ (right) fields against the analytic retarded fields. The convergence order approaches $2$ for late times, while at earlier times the fields are only converging as $1$st order. This is expected and will be explained in the end of this section.

\begin{figure}[H]
  \begin{subfigure}[b]{0.48\textwidth}
    \includegraphics[width=\textwidth]{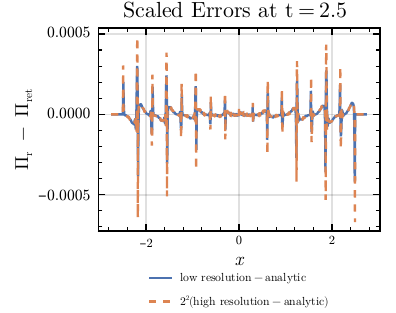}
    \label{fig:pifield-errors-retarded-high-res}
  \end{subfigure}
  \hfill
  \begin{subfigure}[b]{0.48\textwidth}
    \includegraphics[width=\textwidth]{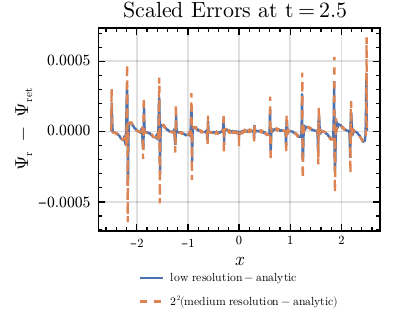}
    \label{fig:psifield-errors-retarded-high-res}
  \end{subfigure}
  \caption{Scaled errors of the numerical fields $\Pi_\mathrm{r}$ (left) and $\Psi_\mathrm{r}$ (right) at $t=2.5$ for resolutions $h_\mathrm{low}=1/512$ and $h_\mathrm{high}=1/1024$, also comparing to the analytic retarded solution. We see an overall second order convergence, overlaid by spikes for first order convergence (which do not affect the overall convergence order as shown in Figure \ref{fig:field-retarded-convergence-order}. The spikes are caused by the slope limiter near extrema of the solution, as explained in the main text, and are a general feature of slope limiters.}
  \label{fig:field-scaled-errors-retarded}
\end{figure}

We further investigate the convergence order of the numerical fields against the analytic retarded solution for late times as $t=2.5$ in Figure \ref{fig:field-scaled-errors-retarded}, by plotting the scaled errors, scaled for 2nd order convergence. We observe the presence of spikes in the errors that do not align on top of each other after scaling for 2nd order. The location of these spikes are at the peaks of the waveforms (see Figure \ref{fig:field-resolutions}) and discontinuities. At a peak the slope changes sign, and as a result the slope limiter is choosing the slope calculation corresponding to an upwind calculation (1st order) instead of a 2nd order slope. This is a well-known artifact of slope-limited schemes when applied to non-smooth data. Away from these regions, the errors align on top of each other after scaling, showing second order convergence as expected for smooth data.

\begin{figure}[H]
  \begin{subfigure}[b]{0.48\textwidth}
    \includegraphics[width=\textwidth]{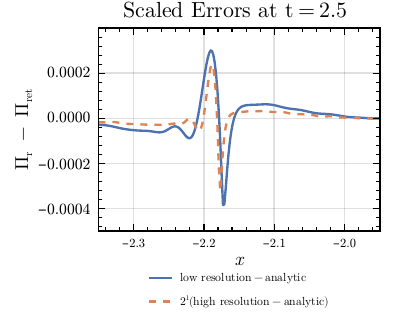}
    \caption{Scaled errors for $\Pi_\mathrm{r}$ for 1st order convergence.}
    \label{fig:pifield-scaled-error-peak-1st}
  \end{subfigure}
  \hfill
  \begin{subfigure}[b]{0.48\textwidth}
    \includegraphics[width=\textwidth]{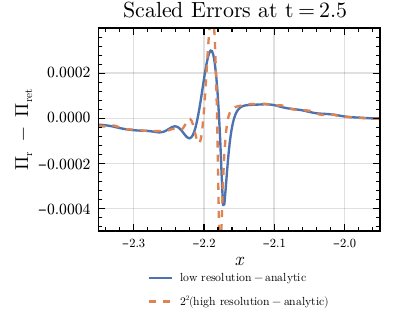}
    \caption{Scaled errors for $\Pi_\mathrm{r}$ for 2nd order convergence.}
    \label{fig:pifield-scaled-error-peak-2nd}
  \end{subfigure}
  \caption{Scaled errors for $\Pi_\mathrm{r}$ scaled for 1st (left) and 2nd (right) order near a spike location.}
  \label{fig:pifield-scaled-error-peak}
\end{figure}

To illustrate this more explicitly we zoom in on one particular spike for $\Pi_\mathrm{r}$ at $r=2.5$ and show the errors, scaled for both 1st and 2nd order convergence, in figure \ref{fig:pifield-scaled-error-peak}. We see the errors generally align on top of each other for the smooth region in the 2nd order (right) subfigure \ref{fig:pifield-scaled-error-peak-2nd}, while the peak is not aligning well. The scaled high resolution error overshoots, indicating a lower convergence order there. The 1st order (left) subfigure \ref{fig:pifield-scaled-error-peak-1st} shows that the peak aligns reasonably well with a scaling for a 1st order error.

\begin{figure}[H]
  \begin{subfigure}[b]{0.48\textwidth}
    \includegraphics[width=\textwidth]{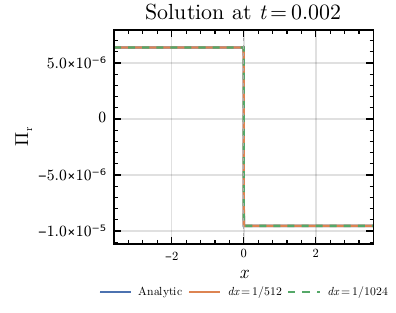}
  \end{subfigure}
  \hfill
  \begin{subfigure}[b]{0.48\textwidth}
    \includegraphics[width=\textwidth]{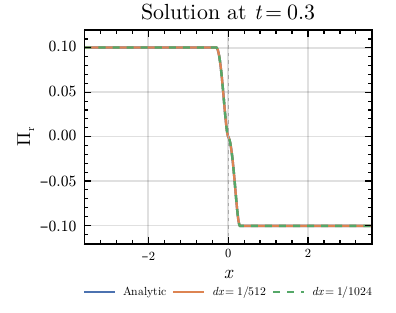}
  \end{subfigure}
  \caption{Resolutions for $\Pi_\mathrm{r}$ at $t=0.002$ and a bit later at $t=0.3$.}
  \label{fig:pifield-resolutions-early-time}
\end{figure}

Having understood the effect of the slope limiters and how they affect the convergence order in non-smooth regions and near extrema, we now explain why the overall convergence order is reduced at early times, as promised when discussing Figure \ref{fig:field-retarded-convergence-order}. The solution starts with the initial conditions $\Pi_\mathrm{r}=\Psi_\mathrm{r}=0$. At the first time step, the flux terms will be zero, and the source terms will be added to the RHS as is. These are discontinuous for a 1D particle field. In the next time step, this non-smoothness makes the slope limiters limit the slopes, reducing the convergence order. Figure \ref{fig:pifield-resolutions-early-time} shows the solution for the remainder field $\Pi_\mathrm{r}$ at a very early time where the discontinuity at the particle location is clearly visible. At a later time $t=0.3$, $\Pi_\mathrm{r}$ has a non-trivial shape for $|x| \leq t$, and the discontinuity at the particle does not dominate the numerical error any more. For more extensive discussion about high-resolution FVM with slope limiters, we refer the reader to \cite{LeVeque2002}.

Overall, the finite volume discretization is working well, in line with the expected behavior for discontinuous source terms and slope limiters. This implementation of our method converges overall to second-order as expected. The trajectory of the particle is given a priori, i.e. the particle does not lose energy or momentum while radiating. The focus of this subsection is to demonstrate that the non-regular source terms can be handled well by numerical methods.

Unfortunately, our attempts to let the particle be accelerated by the gradient of the remainder field failed, likely because our (ultimately piecewise constant) FVM discretization did not provide sufficiently accurate approximation of the field gradient near the particle. Further work is required to make self-consistent particle/field calculations work with finite volume discretizations.

We show a fully interacting particle/field system in the next subsection.

\subsection{Spectral method results}

We choose a domain $[-L, L]$ with $L=1$. We expand the fields
$\Phi_\mathrm{r}$ and $\Pi_\mathrm{r}$ into Fourier modes and impose Dirichlet boundary
conditions $\Phi_\mathrm{r} = \Pi_\mathrm{r} = 0$. We choose basis functions $b_n(x) =
  \sin \left( \pi n (x+1)/2 \right)$ which strongly impose these
boundary conditions. To avoid contamination of our results by boundary
effects, we restrict our analysis to the time interval $[0, 0.4]$ and
consider only the spatial region $[-0.5, 0.5]$ when evaluating the
results. We choose a particle with $Q=1$ and $M_0=1$.

We also use a non-relativistic approximation, i.e. we remove all
relativistic terms from the equations of motion. We do so to be able
to better compare to analytic results. The heart of our method is not
affected by using Newtonian physics.

We choose an oscillation period $P$ of $1/10$.
We consider two different oscillation amplitudes $A$ of $P/10$
and $P/100$. We use four different resolutions $N$ with $200$, $400$,
$800$, and $1600$ expansion coefficients.

\begin{figure}[H]
  \includegraphics[width=\textwidth]{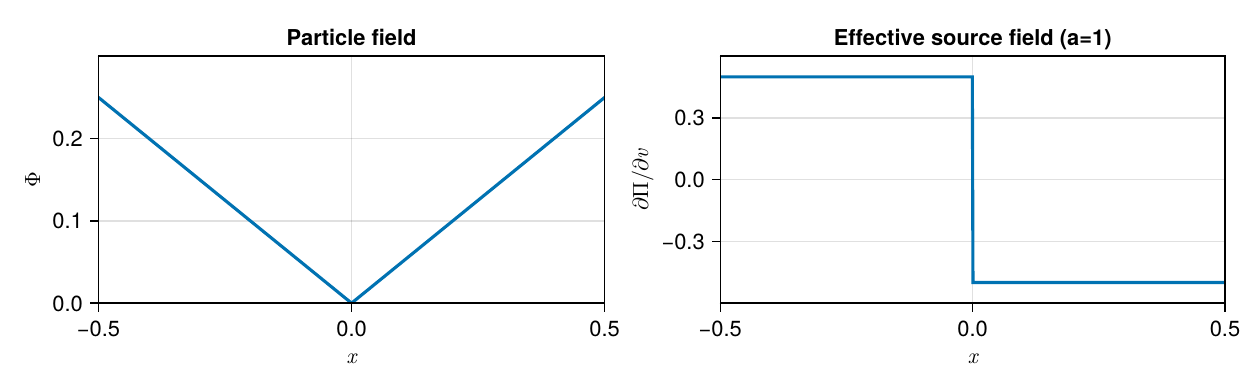}
  \caption{\label{fig:self-field}A particle's self-field and its
    effective source. In one space dimension, a particle's field is
    proportional to $|x|$ instead of $1/r$. The radiation field of a
    particle is given by its effective source
    $\partial\Pi_\mathrm{lab} / \partial v$, multiplied by the
    particle's acceleration $\partial_t v_\mathrm{p}$. (A non-accelerating
    particle does not radiate, and its effective source is zero.)
    (We show the effective source for $a=1$.) This Figure neglects
    relativistic effects such as the Lorentz contraction.}
\end{figure}

We illustrate a particle's self-field and its effective source in
Figure \ref{fig:self-field}. As mentioned earlier, the particle's
field $\Phi$ grows linearly with distance. $\Phi$ corresponds to the
potential $U$ in electrodynamics. $\Pi$ (not shown here), which
corresponds to the electric field $E$ in electrodynamics, is constant
in space. The effective source changes over a length scale given by
the particle size $\ell$, i.e. it ``looks'' discontinuous.

These properties hold only in one spatial dimension. In three
dimsions, the particle field falls off as $1/r$, and the effective
source would fall off as $1/r^2$ near the particle. (Even in the limit
$\ell \to 0$, the particle field and its effective source remain
integrable in a neighbourhood of the particle.)

\begin{figure}[H]
  \includegraphics[width=\textwidth]{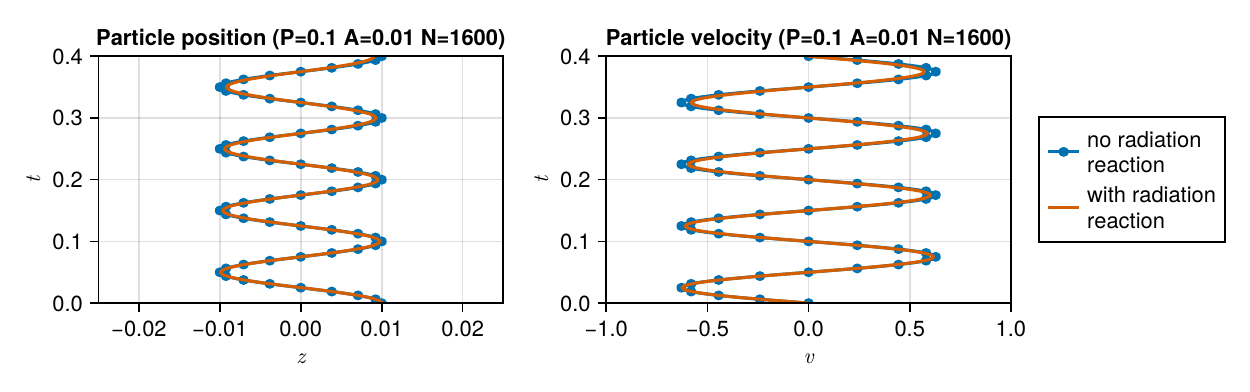}
  \caption{\label{fig:pos-vel}Particle position and velocity vs.
    time both without and with radiation reaction. Both graphs are
    very similar. The self-interacting particle's amplitude and
    maximum velocity decrease slightly over time because the
    particle loses energy as it radiates.}
\end{figure}

\begin{figure}[H]
  \includegraphics[width=\textwidth]{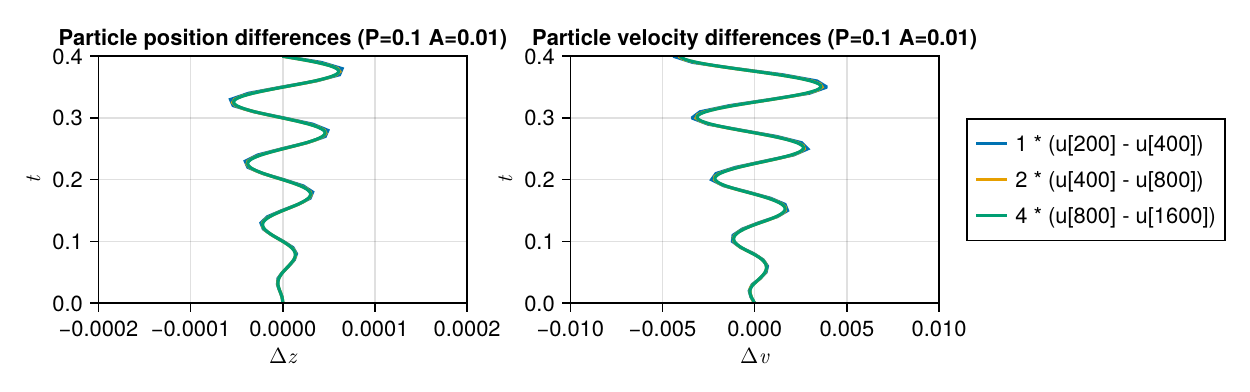}
  \caption{\label{fig:pos-vel-diff}A convergence test showing the
    scaled particle position and velocity differences for four
    resolutions. As expected our discretization is first-order
    convergent. This low convergence order is expected for a
    spectral representation of the discontinuous effective source.}
\end{figure}

We show position and velocity of the oscillating particle in Figure
\ref{fig:pos-vel} and show a respective convergence test in Figure
\ref{fig:pos-vel-diff}. We compare two cases -- a particle where the
radiation reaction has been turned off (the particle sees only the
harmonic potential, i.e. it moves on a pre-specified trajectory) and a
particle that sees the radiation field it generates, and which thus
loses energy. This energy loss is visible as a reduction in the
oscillation amplitide and peak velocity. The convergence tests shows
the rescaled error in the position of the radiating particle,
demonstrating first order convergence (as expected).

\begin{figure}[H]
  \includegraphics[width=\textwidth]{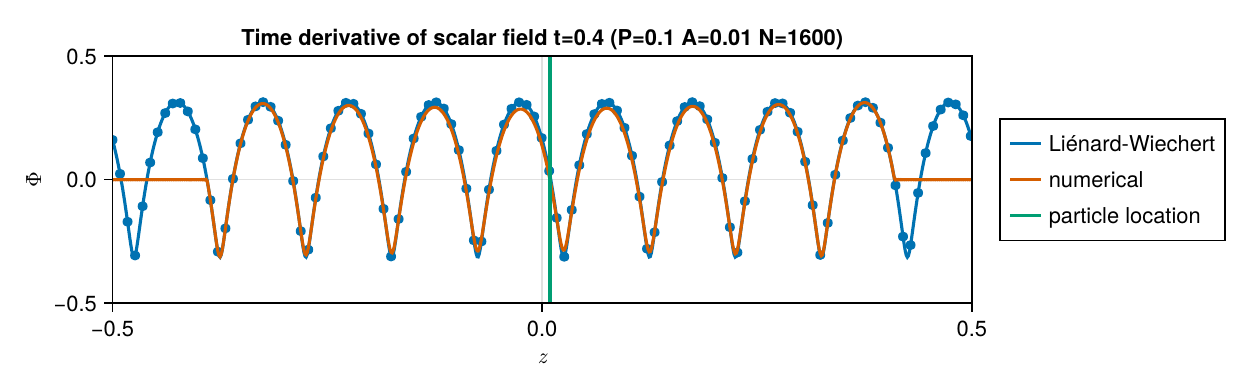}
  \caption{\label{fig:field}Generated radiation field at the final
    time, compared to the Liénard-Wiechert approximation which
    ignores the radiation reaction. The numerical calculation
    initially agrees well with the Liénard-Wiechert approximation
    (near $x=\pm0.4$), but later produces less radiation because the
    oscillating particle has lost energy. The radiation field is
    lopsided because the particle oscillates fast ($v>0.5$), leading
    to a significant Doppler effect.}
\end{figure}

\begin{figure}[H]
  \includegraphics[width=\textwidth]{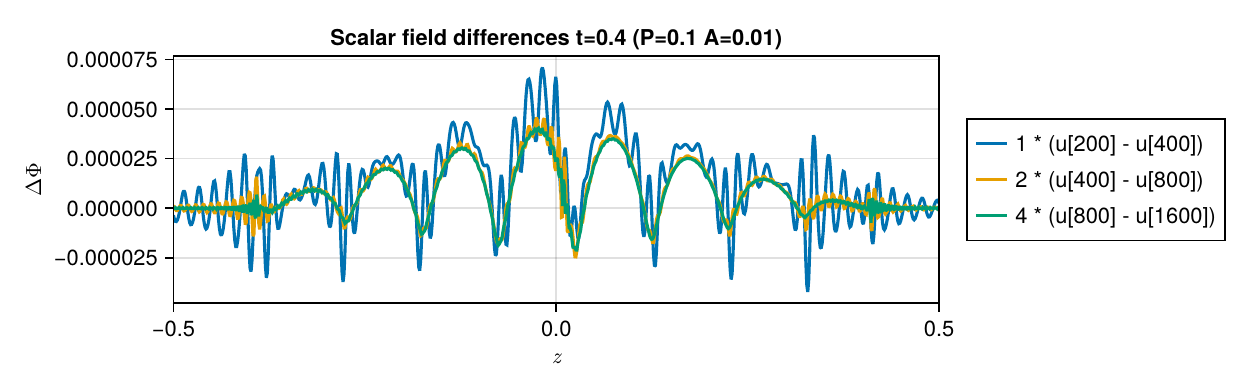}
  \caption{\label{fig:field-diff}A convergence test showing the
    scaled radiation field differences at the final time. As
    expected, our discretization is first-order convergent. This low
    convergence order is expected for a spectral representation of
    the discontinuous effective source.}
\end{figure}

We show the radiation field $\Pi$ in Figure \ref{fig:field} and a
respective convergence test in Figure \ref{fig:field-diff}. We
subtract the initial field, i.e. $\Pi(t=0)$, to clarify the
presentation. This shows the radiation field at a late time ($t=0.4$);
at this time, the particle is located at $z=0.01$, where the radiation
field crosses zero.

The waves to the right of the particle propagate to the right, the
waves on the left propagate to the left. The radiation field has in
general a discontinuity at the particle location (although there is no
discontinuity at the show time $t=0.4$).

The field has a clearly visible asymmetry -- the maxima and minima of
the field have a visibly different shape. This is caused by the
Doppler effect; when the particle is moving to the right, the
right-moving wave is squeezed, and when the particle is moving to the
left, the right-moving wave is stretched. The maximum particle
velocity is about $0.6$, which is a significant fraction of the wave
speed $1$, although the simulation is non-relativistic.

The Liénard-Wiechert approximation assumes that
the particle is on a force trajectory, i.e. it ignores the self-force.
At early times (i.e. near $|z|=0.4$), it agrees well with our
numerical calculation. At late times (i.e. closer to the origin), our
numerical calculations produce less radiation because the particle has
lost energy and its oscillation amplitude has shrunk. Outside of
$|z|=0.4$ the radiation field is zero, because the radiation has not
yet reached this region.

We see first-order convergence of the radiation field as well.

\begin{figure}[H]
  \includegraphics[width=\textwidth]{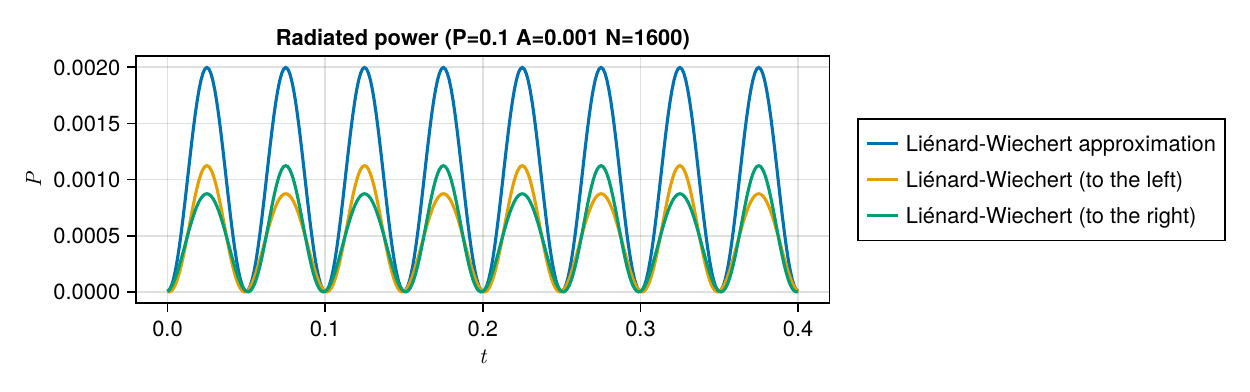}
  \caption{\label{fig:power}Particle estimated radiation power vs.
    time, calculated via a Liénard-Wiechert potential.}
\end{figure}

\begin{figure}[H]
  \includegraphics[width=\textwidth]{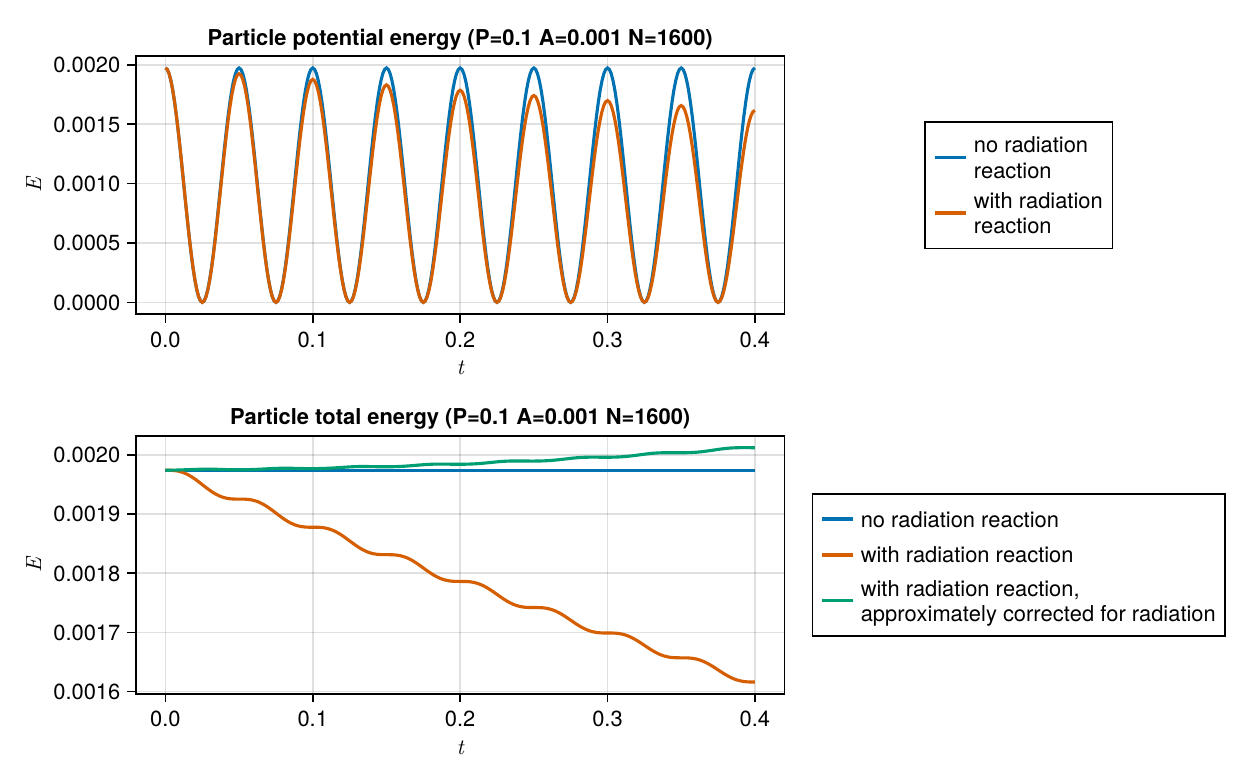}
  \caption{\label{fig:energy}Particle energy vs. time. The particle
    energy decreases over time. It does so in bursts because the
    particle is oscillating forth and back and not in a circular
    motion. The particle's acceleration is largest when it reverses
    direction and is zero in between, and the radiated power is
    proportional to the particle's acceleration. This is well
    modeled by the (averaged) expected radiation power calculated
    via a Liénard-Wiechert potential.
    At late times the Liénard-Wiechert approximation overestimates
    the radiated power because the particle has lost energy.}
\end{figure}

We examine the particle's radiative energy loss in Figures
\ref{fig:power} and \ref{fig:energy} as function of time.

In Figure \ref{fig:power}, we estimate the particles radited power via
a Liénard-Wiechert approxmation. We show the left-propagating, the
right-propagating, as well as the total emitted power. This shows that
the particle radiates in bursts; it radiates most when it accelerates
most (when it is near the extremum of its oscillation), and does not
radiate when it is at the origin. The left/right asymmetry, caused by
the Doppler effect, is also clearly visible.

In Figure \ref{fig:energy} we show the potential energy of the
particle for two cases, a non-self-interacting and a self-interacting
particle. The self-interacting particle loses energy to radiation; it
has lost about 20\% of its energy at the end of the simulation.

We also show the total energy of the particle for both cases. The
total energy is the sum of the potential and kinetic energy. The
non-self-interacting particle's energy is constant in time, the
self-interacting particle's energy decreases monotonically over time.
The energy loss is not linear because the particle only radiates when
it accelerates in its oscillatory motion. We estimate the energy loss
due to radiation via a Liénard-Wiechert approxmation (``approximately
corrected for radiation''). This approximation is not exact because it
over-estimates the radiation at late times, it does not account for
the reduction in oscillation amplitude over time.

\section{Summary}

We have presented a new approach for modeling small (unresolved) particles in numerical calculations. Our approach treats particles as small but finite-sized, and models the particle field effectively as an additional basis function on top of a regular remainder field. This sidesteps any issues arising from handling infinities, and subsumes accuracy and convergence into well-known mechanisms for numerical methods.

We showed numerical calculations using two different discretization methods (finite volumes and spectral methods) with excellent agreement to analytic solutions. The fully self-interacting case agrees well with analytic approximations. We emphasize that it is necessary to ensure that the effective sources created by the particle acceleration are properly projected into the discrete function space. While this can be done analytically, we show that this can also be done numerically, which may be preferable because the source terms are complicated.

While our example calculations were performed in one dimension, our approach generalizes to any number of dimensions. In three dimensions, the effective source terms fall off as $1/r^2$ but still contain directional discontinuities. To improve convergence one could move to more advanced numerical methods that are better suited to handle such sources. One possible approach would be a finite element method where the element containing the particle is split into sub-elements so that the particle can be placed into element corners. Different from our one-dimensional calculations, where the solution of the Poisson equation grows as $|x|$, outer boundaries will not be an issue in three dimensions.

To limit the region affected by the effective sources, one can employ a window function in the definition of the particle field $\Phi_\mathrm{p}$. This leads to additional terms in the effective sources, but can reduce the overall computational cost. Indeed the sole purpose of such a window function is to reduce computational cost; a particle field $\Phi_\mathrm{p}$ with infinite support is fully self-consistent.

Scalar particles and electrodynamic particles are described by their charge, mass, position, and velocity. Particles describing a compact object in relativity have an additional degree of freedom, namely their angular momentum. This could be implemented as a boosted Kerr black hole in some coordinates, or by an ansatz that only models the leading order terms of the singularity. In our approach, only the combined particle and remainder fields have a physical meaning, and the particle field by itself does not need to satisfy any constraints or gauge conditions.

Our approach to limit the physical system to a finite resolution $h \gneq 0$ makes sense for the physical scenarios we described in the introduction, e.g. a small compact object with mass $m$ orbiting a large black hole of mass $M$. The practically interesting range for the resolution $h$ is $M \gg h \gg m$, with the numerical resolution $h$ sandwiched between the sizes of the large and small objects. Our method is designed for this, and describes the small object as particle of size $\ell > 0$. In the limit $h \to 0$, the resolution is at some point finer than the small object, and we then have $M \gg m \gg h$. The small object can be fully resolved, and case we do not need to treat it as a particle any more. Thus we never need to take the limit $\ell \to 0$. This idea is at the core of our approach.

\begin{acknowledgments}
  We thank Leor Barack and Eric Poisson for useful discussions. SV is grateful for the hospitality of Perimeter Institute where part of this work was carried out. Research at Perimeter Institute is supported in part by the Government of Canada through the Department of Innovation, Science and Economic Development and by the Province of Ontario through the Ministry of Colleges, Universities, Research Excellence and Security. We acknowledge the support of the Natural Sciences and Engineering Research Council of Canada (NSERC). Nous remercions le Conseil de recherches en sciences naturelles et en génie du Canada (CRSNG) de son soutien.
\end{acknowledgments}

\bibstyle{apsrev4-2}
\bibliography{paper}

\end{document}